\documentclass[aps,prd,twocolumn,superscriptaddress,nofootinbib]{revtex4-2}

\usepackage{graphicx}
\usepackage{xspace}
\usepackage{xcolor}
\usepackage{hyperref}
\hypersetup{
    colorlinks=true,
    linkcolor=blue,
    filecolor=magenta,      
    urlcolor=cyan,
}
\usepackage{booktabs}
\usepackage{amssymb}
\usepackage{siunitx}
\usepackage{amsmath}
\usepackage{breqn}

\begin{document}

\preprint{LIGO-P1900360}

\title{The failure of the Fisher Matrix when including tidal terms: Considering construction of template banks of tidally deformed binary neutron stars}

\newcommand{\icg}{\affiliation{University of Portsmouth, Institute of Cosmology and Gravitation, Portsmouth PO1 3FX, United Kingdom}}
\author{Ian Harry}
\icg
\author{Andrew Lundgren}
\icg

\date{\today} 

\begin{abstract}
Recent gravitational-wave observations have begun to constrain the internal physics of neutron stars. However, current detection searches for neutron star systems assume that potential neutron stars are low-mass black holes, ignoring any affect on the gravitational-wave signal due to the internal neutron-star physics. We wish to create a template bank of binary neutron star waveforms including the effect of tidal deformability. However, we find that the Fisher matrix, which is commonly used to approximate match calculations when placing template banks, is unsuitable to predict the match between two binary neutron star waveforms. We find that the Fisher matrix can predict errors on the mismatch that are larger than $100\%$ when attempting to identify waveforms with a match of $0.97$. We explore the regime in which the Fisher matrix cannot be trusted and examine why it breaks down. We demonstrate that including higher-order terms in the Taylor series expansion of the match can reliably compute matches for these examples, but that it is prohibitively computationally expensive to do so. Finally, we demonstrate that stochastic placement can still be used to construct a template bank of tidally deformed neutron-star waveforms.
\end{abstract}


\maketitle

\section{\label{sec:intro}Introduction}

Second generation gravitational-wave observatories, Advanced LIGO~\cite{TheLIGOScientific:2014jea} and Advanced Virgo~\cite{TheVirgo:2014hva}, have recently opened the gravitational-wave Universe to us.
Observations of binary black holes~\cite{Abbott:2016blz,LIGOScientific:2018mvr}, binary neutron stars~\cite{TheLIGOScientific:2017qsa,Abbott:2020uma} and neutron-star--black hole mergers~\cite{Abbott:2020khf} have demonstrated that a rich wealth of systems can be observed through gravitational waves.
As these observatories continue to increase in sensitivity, and are joined by additional observatories in Japan~\cite{Akutsu:2020his} and India~\cite{LIGOIndia} we can continue to expect new discoveries and better understanding of the gravitational-wave Universe in the coming years.

The observation of compact binary mergers has required complex search algorithms~\cite{Allen:2005fk, Babak:2012zx, Hooper:2011rb, Usman:2015kfa, Adams:2015ulm, Messick:2016aqy, Venumadhav:2019tad}.
These search algorithms rely on matched-filtering a large set of filter waveforms against the full set of data~\cite{Allen:2005fk, Babak:2012zx}.
The filter waveforms are generated from models designed to create reliable waveforms for any input physical parameters~\cite{Bohe:2016gbl, Khan:2015jqa}.
These sets of filter waveforms, commonly referred to as a ``template bank'', are normally constructed such that any physical signal in the parameter space of interest would be found with no more than a fixed loss (normally 3\%) in signal-to-noise ratio~\cite{Babak:2012zx}.

There are two main methods that are used for creation of template banks.
The first approach, stochastic placement, involves randomly choosing a very large set of points within the parameter space and then iterating through these points accepting only points that are not ``close'' to any point already accepted~\cite{Harry:2008yn, Babak:2008rb, Harry:2009ea, Manca:2009xw, Ajith:2012mn, Privitera:2013xza, Capano:2016dsf}.
The second approach, geometric placement, involves using the Fisher matrix to define a metric on the parameter space and then placing a lattice of points to cover the full space~\cite{Sathyaprakash:1991mt, Poisson:1995ef, Balasubramanian:1995bm, Owen:1995tm, Owen:1998dk, Babak:2006ty, Cokelaer:2007kx, Brown:2012qf, Harry:2013tca}.
The stochastic placement offers the benefit of flexibility, it can be applied to any placement problem.
The geometrical placement offers the benefit that it will place more efficient banks---in terms of fewer templates to achieve a given covering criterion---but does require an underlying metric accurately describing the parameter space.
Current wisdom states that the geometric placement is the best option when placing template banks of binary neutron stars, when analytical metrics describing inspiral-only waveforms can be accurately applied. For heavier systems, where the merger is important, stochastic placement is commonly used. In addition,
recent work has explored the development of hybrid methods attempting to combine the advantages of both methods to place efficient template banks~\cite{Roy:2017qgg, Roy:2017oul}.

LIGO/Virgo searches for compact binary mergers have always searched for binary neutron star systems using waveforms that assume both bodies are point particles~\cite{Babak:2012zx}.
In reality, the internal composition of the neutron star can be important in the dynamics of the system. Tidal-induced deformation~\cite{Damour:1982wm, Flanagan:2007ix, Vines:2011ud, Damour:2012yf, Gralla:2017djj}, spin-induced deformation~\cite{Poisson:1997ha, Bohe:2015ana} and the complex physics describing the post-merger behaviour of a binary neutron star merger (see e.g. \cite{Baiotti:2016qnr}) can all leave observable imprints in the gravitational-wave signal.
In terms of observing gravitational-wave signals from binary neutron star mergers it has already been demonstrated that the tidal-induced deformation does have an impact on the observability of such systems~\cite{Cullen:2017oaz}. 
This effect is not large enough to cause us to miss ``loud'' signals such as GW170817 or GW190425, but can cause some reduction in sensitivity to signals close to the detection threshold~\cite{Cullen:2017oaz}.
From measurements of GW170817 and GW190425 we already have some indication of the equation-of-state governing the internal physics of neutron stars~\cite{Abbott:2018exr, LIGOScientific:2019eut}.
It is natural to include this knowledge in construction of future template banks for LIGO and Virgo searches and place lattices of templates including the neutron stars' equation of state.

In this paper we will aim to demonstrate how to place template banks of binary neutron star waveforms that include the effects of tidal induced and spin induced deformation.
We begin by extending the current geometrical approaches to include the high-order tidal terms in the metric used.
In doing so, we demonstrate the inadequacy of the Fisher Matrix approximation for predicting matches between waveforms with tidal corrections
modelled in the frequency domain using the post-Newtonian approximation.
This result has implications not only in terms of searches, but also in parameter inference studies that use the Fisher Matrix as a proxy for numerical match computations.
We further explore the validity of the Fisher Matrix in more general terms, and show how including higher order terms in the analytical Taylor series expansion of the match can improve the accuracy, but are difficult to include for template placement.
Finally we demonstrate that stochastic methods are, as expected, capable of placing template banks of binary neutron stars including tidal terms, and discuss the efficiency of such methods.

\section{The Fisher Matrix}
\label{sec:fmatrix}

We begin by describing how the Fisher Matrix is currently used to approximate the overlap between two waveforms in gravitational-wave astronomy.
This formalism is widely used in the literature, but we include it here for completeness and for clarity when we later introduce higher-order corrections to the Taylor series expansion of the overlap.
Specifically, we draw from the methods and formalism derived in ~\cite{Sathyaprakash:1991mt, Poisson:1995ef, Balasubramanian:1995bm, Owen:1995tm, Owen:1998dk, Porter:2002vk, Babak:2006ty, Cokelaer:2007kx, Brown:2012qf, Harry:2013tca} when formulating
this section.

A common problem in gravitational-wave astronomy is one where we want to determine the overlap between two waveforms.
This is a measure of how well two waveforms agree with each other, and can be used to assess how well we might observe a given signal if using a search template with different physical parameters as a filter.
It can also be used to determine the chance that detector noise will make it impossible to distinguish two waveforms with different parameters.
This overlap can be computed numerically as:
\begin{equation}
    \left\langle h_1 | h_2 \right\rangle = 4 \Re \left[ \int^\infty_0 df \frac{{\tilde{h}_1}^{*}(f) \tilde{h}_2(f)}{S_h(f)} \right].
\end{equation}
It is standard practice to normalize $h_1$ and $h_2$ such that $\left\langle h_1 | h_1 \right\rangle = 1$ and $\left\langle h_2 | h_2 \right\rangle = 1$.
This implies that if $\left\langle h_1 | h_2 \right\rangle = 1$ then these two waveforms are identical, whereas if $\left\langle h_1 | h_2 \right\rangle = 0$ then the waveforms are completely ``orthogonal'' in parameter space.
Another common measure directly related to the overlap is the ``match'', which is the overlap maximized over a phase and time-shift
between the two waveforms.

There are many purposes, of which template bank placement is one example, where the match between two waveforms must be computed many times.
This can computationally become very expensive~\cite{Harry:2009ea}.
In such cases it is strongly desirable to be able to rapidly compute the match between any two waveforms.
More specifically, we often want to compute the match between any two waveforms \emph{if} it is near to 1, or quickly identify that the match is not close to 1.
Many optimizations exist for the numerical match code, used in the context of template bank placement~\cite{Privitera:2013xza, Capano:2016dsf} or for parameter inference~\cite{Smith:2012du,Canizares:2014fya,Zackay:2018qdy}.

Nevertheless, despite these optimizations, it is much more efficient if the overlap, and match, can be computed analytically.
We define the overlap between two nearby waveforms, one with parameters denoted by $\theta_i$ and another with parameters denoted by $\theta_i + \Delta\theta_i$ as
\begin{equation}
 O\left(\theta_i, \Delta\theta_i\right) = \left\langle h(\theta_i) | h(\theta_i + \Delta\theta_i) \right\rangle.
\end{equation}
By definition $O\left(\theta_i, 0\right) = 1$. If we are interested in computing this quantity in the limit that $\Delta\theta_i$ is small then we can Taylor expand:
\begin{equation}
\label{eq:match_taylor_expand}
    O\left(\theta_i, \Delta\theta_i\right) = O\left(\theta_i, 0\right) + \frac{\partial O}{\partial \theta^i} \Delta \theta_i + \frac{1}{2!} \frac{\partial^2 O}{\partial \theta^i \partial \theta^j} \Delta \theta_i \Delta \theta_j + ...
\end{equation}

$O\left(\theta_i, 0\right)$ must be a maximum, as it is not possible for the overlap to be larger than 1, therefore $\frac{\partial O}{\partial \theta^i}$ must be 0. 
Neglecting higher order terms in the Taylor expansion we
can then write this as
\begin{equation}
\label{eq:fisher_matrix}
 1 - O\left(\theta_i, \Delta\theta_i\right) \approx - \frac{1}{2!} \frac{\partial^2 O}{\partial \theta^i \partial \theta^j} \Delta \theta_i \Delta \theta_j \equiv g_{ij} \Delta \theta_i \Delta \theta_j.
\end{equation}
In doing so we can see that it is natural to define a metric
\begin{equation}
\label{eq:metric_fisher}
g_{ij} = - \frac{1}{2!} \frac{\partial^2 O}{\partial \theta^i \partial \theta^j}
\end{equation}
on the parameter space, known as the Fisher Matrix. Following~\cite{Porter:2002vk} this is often equivalently written as
\begin{equation}
g_{ij} = \frac{1}{2} \left\langle \frac{\partial h}{\partial \theta^i} \middle| \frac{\partial h}{\partial \theta^j} \right\rangle,
\end{equation}
however the form in equation~\ref{eq:metric_fisher} will be more convenient for this work.
The match can then be obtained from the Fisher matrix by projecting out the phase and time-shift coordinates of $g_{ij}$ as
described in~\cite{Owen:1995tm}. In what follows we will use the overlap exclusively, unless we specifically state otherwise.

Generically, the Fisher Matrix here does have a dependence on $\theta_i$---the position in parameter space. However, if $g_{ij}$ can be evaluated quickly the Fisher Matrix offers a way to quickly evaluate the overlap between two waveforms that are ``close'' in parameter space.

\subsection{Fisher Matrix and TaylorF2}
\label{ssec:taylorf2fmat}

The ``TaylorF2'' waveform model~\cite{Droz:1999qx, Buonanno:2009zt}, based on an analytical Fourier transform of the ``TaylorT2'' model~\cite{Blanchet:1996pi, Buonanno:2009zt} via the stationary phase approximation, is a good candidate for use with the Fisher Matrix.
The TaylorF2 waveform model only models the inspiral part of a gravitational-wave signal, using the post-Newtonian expansion~\cite{Droz:1999qx, Buonanno:2009zt}. It can analytically be written as:
\begin{equation}
    \tilde{h}(f) = A(f; \mathcal{M}, D_L, \xi_x) e^{-i \Psi(f; \lambda_{i,j})}.
\end{equation}
The amplitude term here $A$ depends on the chirp mass of the sytem $\mathcal{M}$, the luminosity distance $D_L$ and the orientation angles of the source with respect to the observer $\xi_x$. Most of this dependance is removed when normalizing the waveform. The important evolution here happens in the phase term, $\Psi$. This depends on the intrinsic masses and spins of the component objects via the $\lambda_{i,j}$ coefficients. The phase can be expressed in terms of these coefficients according to
\begin{equation}
 \Psi = 2 \pi f t_c - \phi_c + \sum_i \sum_j \lambda_{i,j} f^{(i-5)/3} \log^j f,
\end{equation}
following the notation used in~\cite{Harry:2013tca}.
These $\lambda_{i,j}$ coefficients are the post-Newtonian coefficients of the waveform. These terms have quite complicated dependencies on the physical parameters, and can be seen in their full form in, for example~\cite{Nitz:2013mxa}. The $i$ coefficient, divided by two, denotes what is commonly referred to when describing the post-Newtonian (PN) orders. For example the $\lambda_{2,0}$ component denotes the ``1PN'' term. The $j$ component denotes the presence, and power, of any log terms in the coefficient.

We also note that the $\phi_c$ and $t_c$ terms, corresponding to the coalescence time and coalescence phase, can also be absorbed into the $\lambda_{i,j}$ coefficients by noting that a shift in $\phi_c$ will correspond to a shift in the $\lambda_{5,0}$ component; the 2.5PN term. Similarly a shift in $t_c$ is equivalent to a shift in the $\lambda_{8,0}$ component; the 4PN term. These terms are then projected out if
one wants to compute the match, instead of the overlap, between two waveforms.

This form is advantageous because, for TaylorF2, it allows one to easily compute the differential of $O(\theta, \Delta \theta_i)$.
In~\cite{Babak:2006ty} an illustration is given of how to compute this differential with respect to any physical parameter.
However, the forms given in~\cite{Babak:2006ty} can be greatly simplified by using the fact that the $\lambda_{i,j}$ coefficients themselves present a natural coordinate system.
Therefore, in this work we compute the Fisher matrix in the
$\lambda_{i,j}$ coordinate system.
Using these coordinates, it can be shown that
\begin{equation}
    \frac{\partial^2 O}{\partial \lambda_{i,j} \partial \lambda_{k,l}} = - 4 \Re \left[ \int^{f_U}_{f_L} df \frac{f^{(i + k - 17) / 3} \log^{j+l} f }{S_h(f)} \right],
\end{equation}
where $f_U$ and $f_L$ are the upper and lower limits to the integration.

In the $\lambda_{i,j}$ coordinate system we can clearly see that the Fisher matrix does not depend on the position in the parameter space, it is globally flat. One can take this further, orthonormalize this parameter space and identify principal directions.
This then offers a natural system in which to place a reduced-dimension lattice of points to form a template bank~\cite{Brown:2012qf} using standard techniques for sphere packing in Cartesian coordinate systems~\cite{Conway:1993}.

In doing this, a number of approximations are made and it is important to clearly state these again before continuing.
Firstly, when using the TaylorF2 waveform model we implicitly assume that it is a fully accurate representation of the emitted gravitational-wave signal.
TaylorF2 does not include the merger and ringdown components of the signal, so will not accurately reproduce the final stages of a binary neutron star merger.
We will not explore this feature in this work, as TaylorF2 is the only waveform for which a globally flat metric has been defined, and the validity of TaylorF2 has been extensively studied elsewhere (e.g ~\cite{Nitz:2013mxa}).
We emphasize that we are in no way dismissing this effect; accurately modelling such waveforms is of paramount importance.
However, we wish to explore in detail the \emph{other} approximations which are normally ignored in the literature.
TaylorF2 is also used extensively in template bank placement, detection searches and parameter estimation studies of binary neutron systems.
The second approximation made is that the TaylorF2 model has some specific start, and end, frequencies outside of which the frequency content is 0.
To achieve a globally flat metric one must also assume that these limits are the same for any value of the physical parameters of the system.
The lower frequency limit is normally chosen through practical constraints, either corresponding to a point at which the power spectral density increases rapidly or limiting the length of the filters used in the search.
The upper frequency cutoff is normally chosen to take some value, between 1000 and 2000Hz, roughly corresponding to the innermost-stable circular orbit of a binary neutron star system where both components have a mass of 1.4$M_{\odot}$.
The power contribution above 1000Hz is small for such systems and so the exact value chosen here, and the fact that it will change with mass, does not matter much. For example for a non-spinning equal-mass binary neutron star system with components of 1.4$M_{\odot}$, 99.7\% of the signal power is emitted in gravitational waves below 1000Hz (assuming the TaylorF2 model).
The third assumption is that terms beyond the Fisher matrix term in the Taylor series expansion of the match are negligible.
In this work, our focus will be on exploring the validity of assuming that the higher-order terms in the Taylor expansion are negligible, in particular for binary neutron-star systems. We also explore the consequences of assuming a constant frequency cutoff.
It is important to highlight that our results are specific to the TaylorF2 waveform,
and the results we find may differ if other waveform models are utilized. 

\section{Neutron stars are not black holes}

There a number of key differences in the gravitational wave signal
emitted by a binary neutron star system and a binary black hole
system with the same component masses and spins. 
These differences arise due to the internal composition of the neutron star and encode information that can let us study the behaviour of matter in one of the most extreme environments in the Universe. 

First, the post-merger phase of a binary neutron star system will be rich with physics (see ~\cite{Abbott:2017dke} for a brief summary).
Will a short-lived hypermassive neutron star be formed, emitting gravitational radiation before collapsing to a black hole~\cite{Baumgarte:1999cq, Shapiro:2000zh, Hotokezaka:2013iia, Ravi:2014gxa}?
Being able to observe the post-merger phase of such a system will offer valuable insights into the internal physics of neutron stars~\cite{Margalit:2017dij, Bauswein:2017vtn, Rezzolla:2017aly, Radice:2017lry}.
However, such signatures will be emitted at high frequencies, which are very challenging for current observatories to detect~(see section 2 in~\cite{Abbott:2018hgk} for a detailed discussion).
Sensitivity to such signatures is a key science objective for proposed third-generation observatories, such as Einstein Telescope and Cosmic Explorer~\cite{Punturo:2010zza,Evans:2016mbw, Reitze:2019iox}.
In this work we neglect such effects.
The TaylorF2 waveform does not include any merger, or post-merger, phase and such effects should not be important until third-generation observatories are built~\cite{Abbott:2018hgk}.

A second effect occurs due to the spherical asymmetry
of rotating neutron stars~\cite{Poisson:1997ha, Bohe:2015ana}.
This effect
adds terms to the post-Newtonian expansion, beginning at 2 PN order (the $\lambda_{4,0}$ term in our notation above).
It has been found that while this term can have a significant effect on the emitted waveform, it is strongly degenerate with the spins of the bodies themselves and does not allow direct measurements of nuclear physics~\cite{Harry:2018hke}.

The most important effect for us to consider is the deformation of the neutron stars due to tidal interactions~\cite{Damour:1982wm, Flanagan:2007ix, Vines:2011ud, Damour:2012yf, Gralla:2017djj}.
This tidal deformation occurs late in the evolution of the system---very near to the merger---with leading terms at 5PN and 6PN order ($\lambda_{10,0}$ and $\lambda_{12,0}$).
As these terms enter the expansion very differently than other terms, they are largely orthogonal, allowing the tidal deformability to be measured~\cite{Hinderer:2009ca}. Therefore it is natural to want to include the effect of these terms in the bank of templates used for observing binary neutron stars. Studies
in~\cite{Cullen:2017oaz,Harry:2018hke} have shown that one incurs a non-negligible
loss in sensitivity if these terms are ignored in detection searches.

\section{Using the Fisher Matrix to predict overlaps with tidal terms}

In this section we begin by exploring the validity of using existing geometric template bank placement methods to place a template bank of neutron-star systems including the deformation due to tidal interactions and the deformation due to the bodies' own rotation. 

The method described in~\cite{Harry:2013tca} already includes the formalism needed to include the deformability due to the bodies' own rotation.
This is simply included by adding the effect to the existing post-Newtonian orders with a suitable range of values encompassing the uncertainty in modelling the neutron star's internal physics.
Similarly the tidal deformability can also be included by introducing the 5- and 6-PN terms into the existing set of parameters being considered.
One then allows a suitable range of values, again given by some range of possible equations of state, and proceeds using the methods described in~\cite{Harry:2013tca}.

However, for this approach to work well, the Fisher matrix
approximation must be able to predict well the overlaps between two waveforms that vary
in the 5- and 6-PN terms ($\lambda_{10,0}$ and $\lambda_{12,0}$) corresponding to
changes in the tidal deformation parameters.
To check if this is the case we generate two TaylorF2 waveforms and add a small perturbation in the 5PN and 6PN terms to one of them before computing the overlap.
By numerically varying the ratio of the perturbation in the two terms, and the magnitude of the deviation, we can identify the line of perturbations that produce a specific overlap.
We then use the Fisher Matrix to predict the perturbations needed to produce the same overlap and can then compare the agreement between the predicted Fisher matrix prediction, and the numerical value.
In these overlap calculations we use a lower frequency cutoff
of 15Hz and an upper frequency cutoff of 2048Hz. The waveform model is generated at all values between these limits even if the innermost-stable circular orbit occurs below 2048Hz.

\begin{figure}[tp]
    \includegraphics[width=\columnwidth]{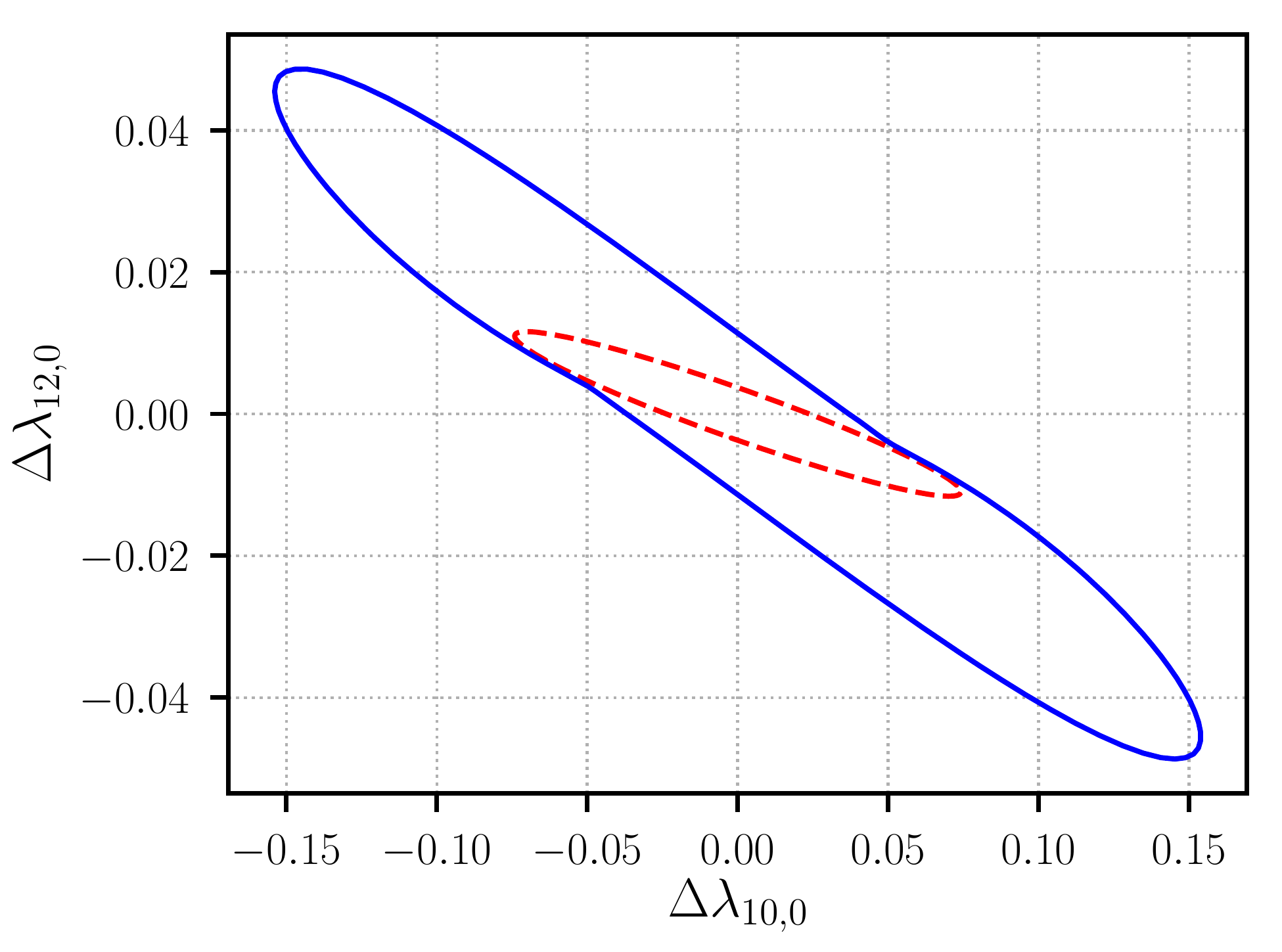}
    \caption{The perturbation that needs to be added to the 5- and
    6-PN terms to achieve an overlap of 0.97 with an unperturbed
    TaylorF2 waveform. The blue solid line shows the numerical overlap.
    The red dashed line shows the overlap predicted by the Fisher matrix.
    For an animated version showing more detail \href{https://icg-gravwaves.github.io/ian_harry/tidal_template_bank/Figure1.mp4}{please see the video here.}
    }
    \label{fig:tidal_metric}
\end{figure}

\begin{figure}[tp]
    \includegraphics[width=\columnwidth]{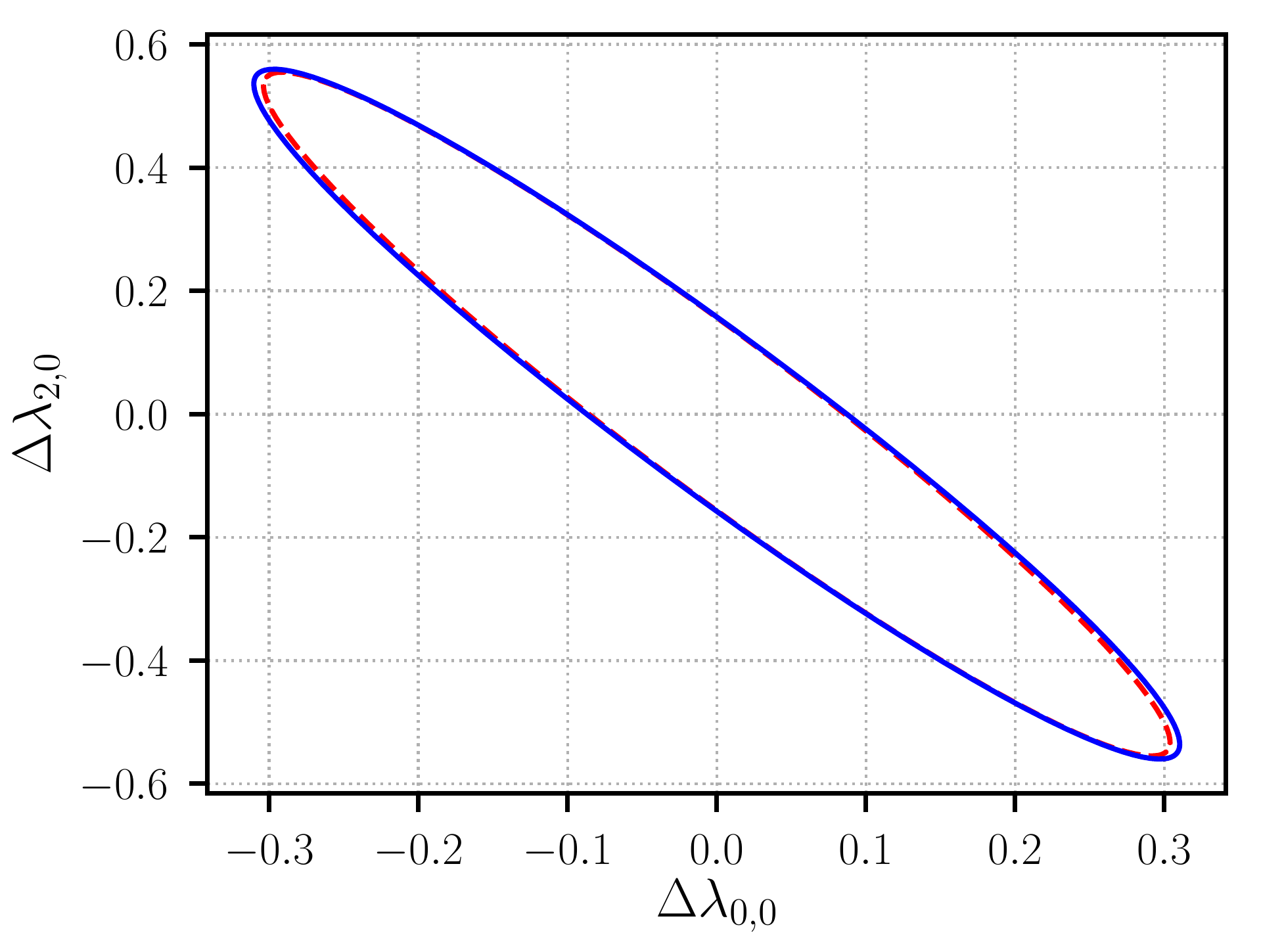}
    \includegraphics[width=\columnwidth]{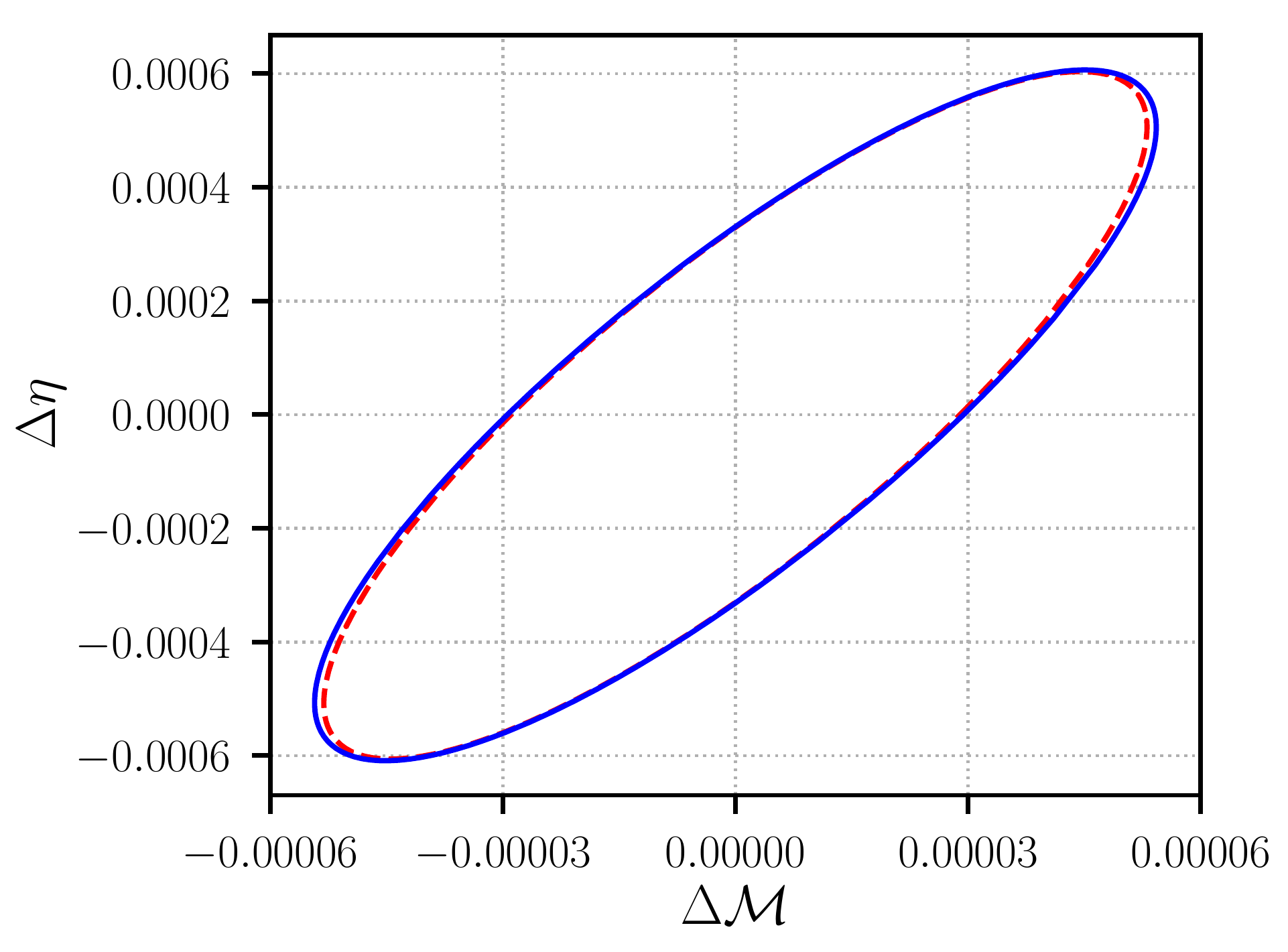}
    \caption{The perturbation that needs to added to the 0- and
    1-PN terms (top) or $\mathcal{M}$ and $\eta$ terms (bottom)
    to achieve a overlap of 0.97 with an unperturbed
    TaylorF2 waveform. The blue solid line shows the numerical overlap.
    The red dashed line shows the overlap predicted by the Fisher matrix.
    For an animated version showing more detail \href{https://icg-gravwaves.github.io/ian_harry/tidal_template_bank/Figure2.mp4}{please see the video here.}
    }
    \label{fig:mass_metric}
\end{figure}

We plot the results of doing this for an overlap of 0.97 in Figure~\ref{fig:tidal_metric}. One can clearly see that the
Fisher matrix is very poorly predicting the overlap between
these two waveforms. For comparison, in Figure~\ref{fig:mass_metric} we produce the same plot, except
that we perturb the 0- and 1-PN terms ($\lambda_{0,0}$ and $\lambda_{2,0}$). In this case the Fisher matrix predicts the
overlap well. We also show in Figure~\ref{fig:mass_metric} the
result of perturbing the physical chirp mass ($\mathcal{M}$) and symmetric mass ratio ($\eta$) terms. Perturbing these terms will
perturb all PN terms up to 3.5-PN order, but the dominant effect will be changes in the leading order PN terms. The Fisher matrix also predicts the overlap well in this case.

It is clear from these plots that the Fisher Matrix can predict well overlaps that are 0.97 or larger when we are changing the
dominant 0- and 1-PN order terms. It also predicts the overlap
very badly when changing the 5- and 6-PN terms where tidal
deformability is first evident. We will now proceed to try to
understand at what PN orders the Fisher matrix performs well,
and where it doesn't. We will also try to understand \emph{why} it works well at some orders, and poorly at others.

\subsection{Where does the Fisher Matrix fail?}

We have demonstrated that the Fisher Matrix fails to correctly predict overlaps when one is varying tidal terms.
This in turn will render the placement algorithm described in~\cite{Harry:2013tca} ineffectual for placing template banks of binary neutron star systems with tidal deformations.
However, it is informative to ask where in the PN expansion the Fisher matrix approximation breaks down and to see if we can understand the conditions under which the Fisher matrix breaks down.

\begin{figure}[tp]
    \includegraphics[width=\columnwidth]{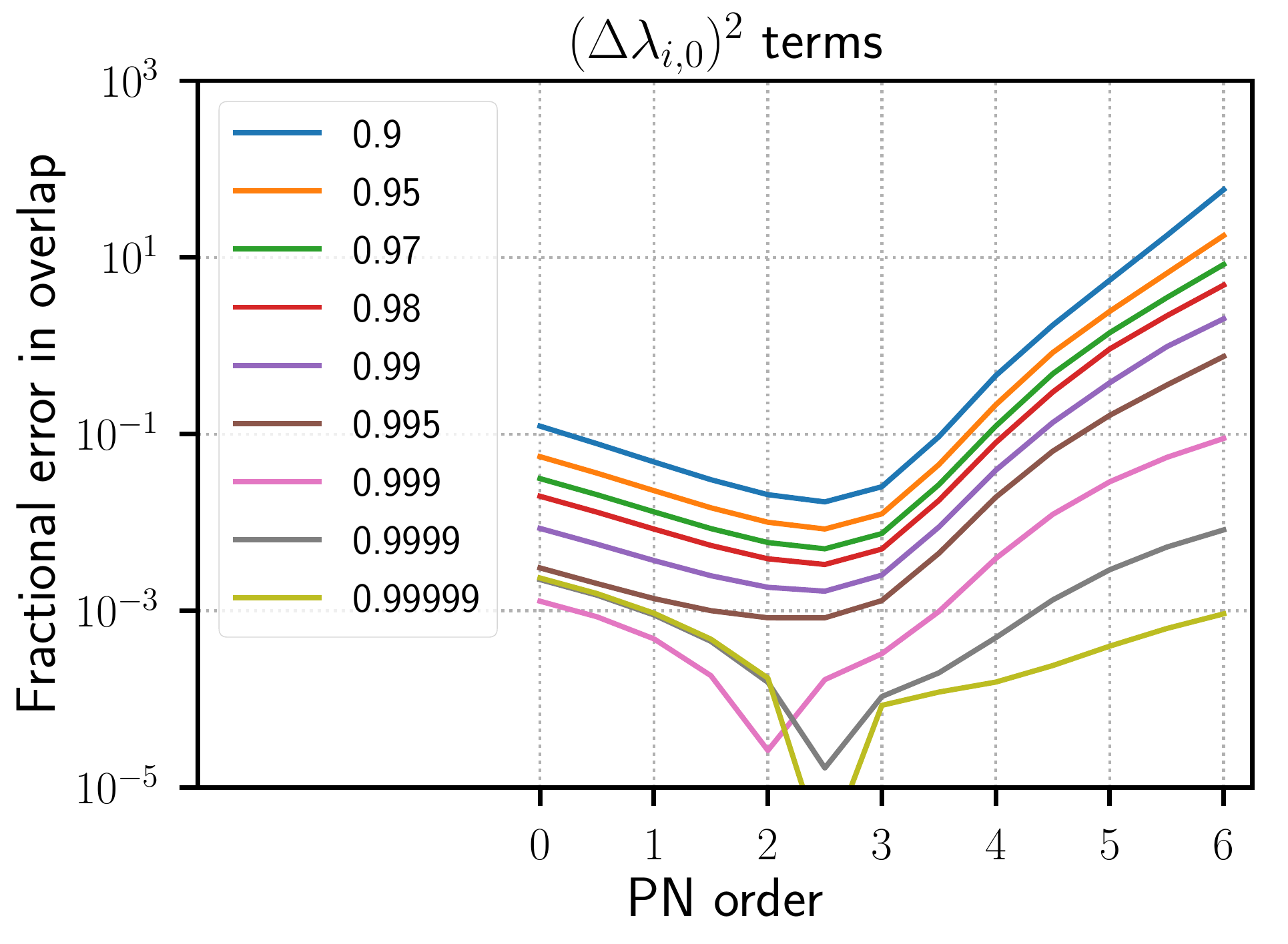}
    \includegraphics[width=\columnwidth]{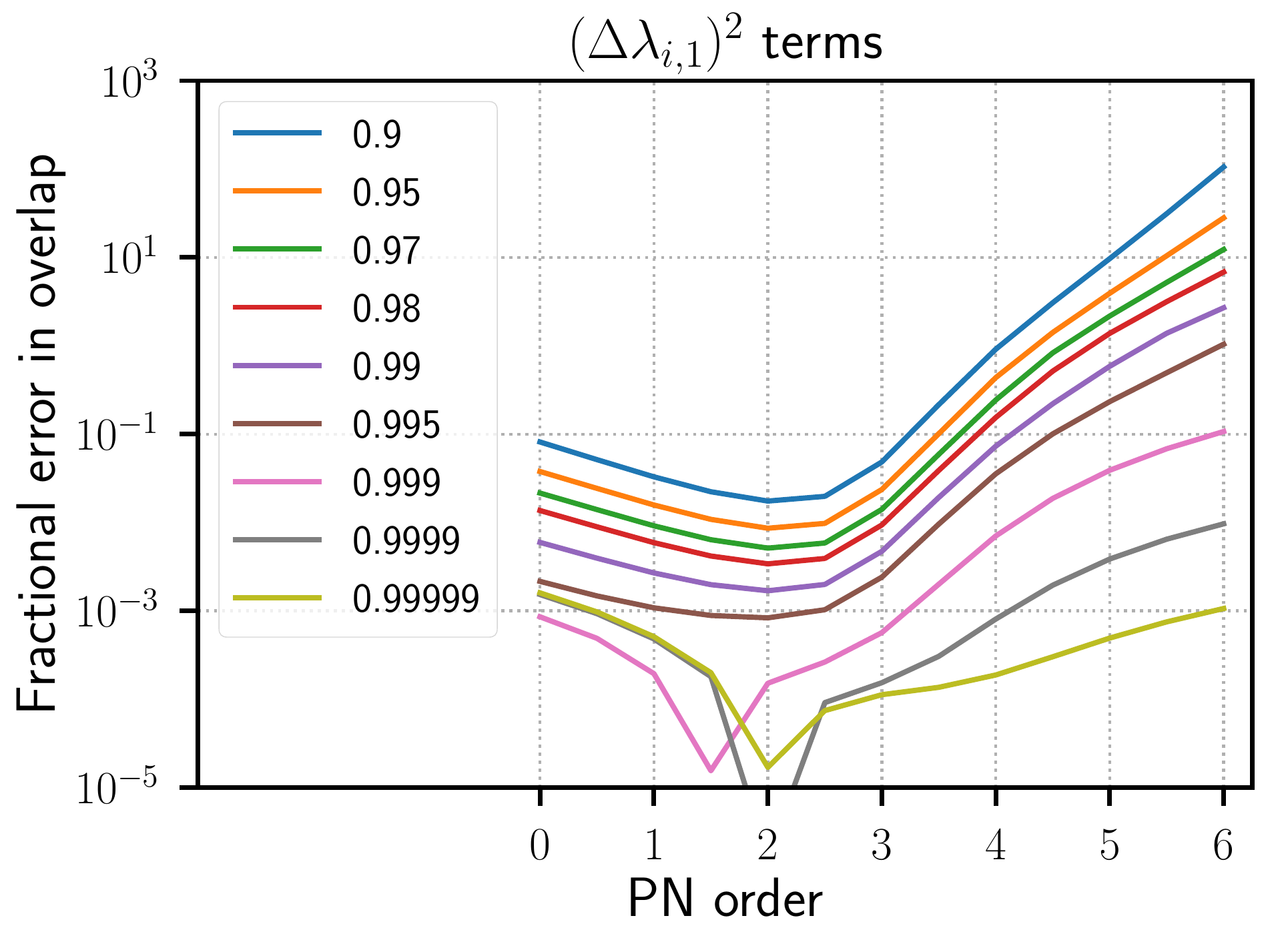}
    \caption{The accuracy at which the Fisher matrix predicts
    the overlap when considering two waveforms
    that differ by a change in a single $\lambda_{i,j}$ term in the
    post-Newtonian expansion. Shown for $\lambda_{i,0}$ (top) and $\lambda_{i,1}$ (bottom).
    }
    \label{fig:change_one_term_fisher}
\end{figure}

We consider the following:
For each of the terms in the post-Newtonian expansion, we vary the value of that term, computing overlaps numerically, until we empirically determine the change needed to produce a specified match (e.g. 0.97) if \emph{only} that term is changing.
We can then compute the overlap that the Fisher Matrix predicts for this change and compare the two.
This can be applied to all terms in the expansion; for example, while there is no 0.5PN term predicted by general relativity, we can still shift the TaylorF2 waveform phasing as if this term was present to determine the mismatch due to its variation.
We do this for all linear order terms between 0 and 6 PN ($\lambda_{0,0}$ to $\lambda_{12,0}$ in our notation) and for all log terms between 0 and 6 PN ($\lambda_{0,1}$ to $\lambda_{12,1}$ in our notation).
This is shown in Figure~\ref{fig:change_one_term_fisher}.
For completeness, the ``fractional error in overlap'' quoted in these plots is defined as:
\begin{equation}
    \frac{(1 - N) - (1 - A)}{(1 - N)},
\end{equation}
where $N$ denotes the numerical overlap and $A$ denotes the analytical prediction, in this case computed using the Fisher matrix.

From Figure~\ref{fig:change_one_term_fisher} we can see that the Fisher matrix appears to be most reliable when computing overlaps at a post-Newtonian order of 2.5 and diverges on either side of this.
For larger post-Newtonian orders this divergence happens quickly and we see again that at post-Newtonian orders associated with tidal effects the Fisher-matrix is not predicting the overlap accurately considering waveforms with a numerical overlap of 0.97.
We notice that when using overlaps very close to 1, the Fisher matrix does predict the overlaps well, but there does appear to be an error of roughly 0.1\% at all post-Newtonian orders except 2.5.
We have not been able to identify where this error comes from---we suspect some subtle systematic effect---but it does not affect our conclusions.

\subsection{The importance of beyond-leading order terms in the match calculation}

\begin{figure}[tp]
    \includegraphics[width=\columnwidth]{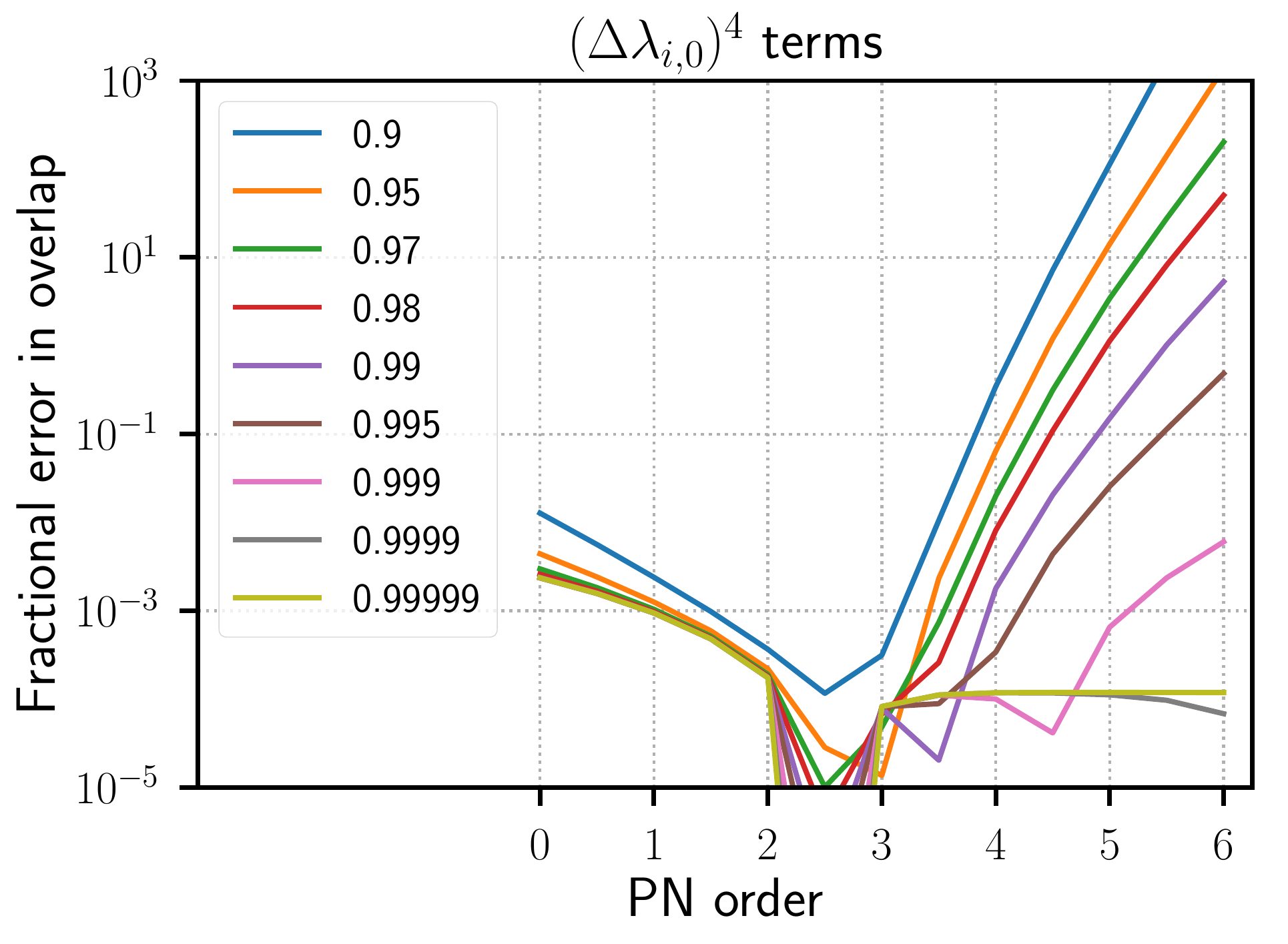}
    \includegraphics[width=\columnwidth]{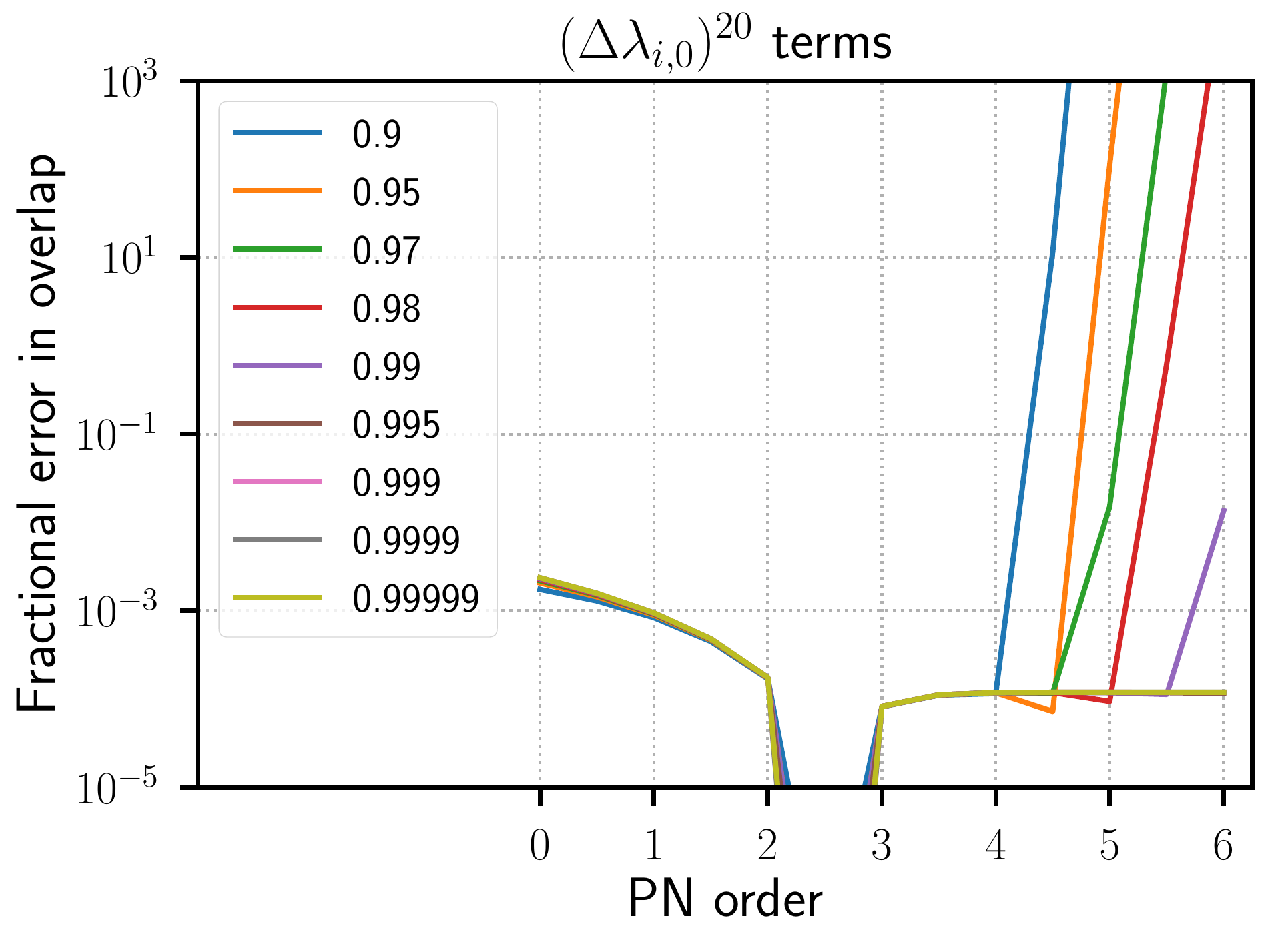}
    \includegraphics[width=\columnwidth]{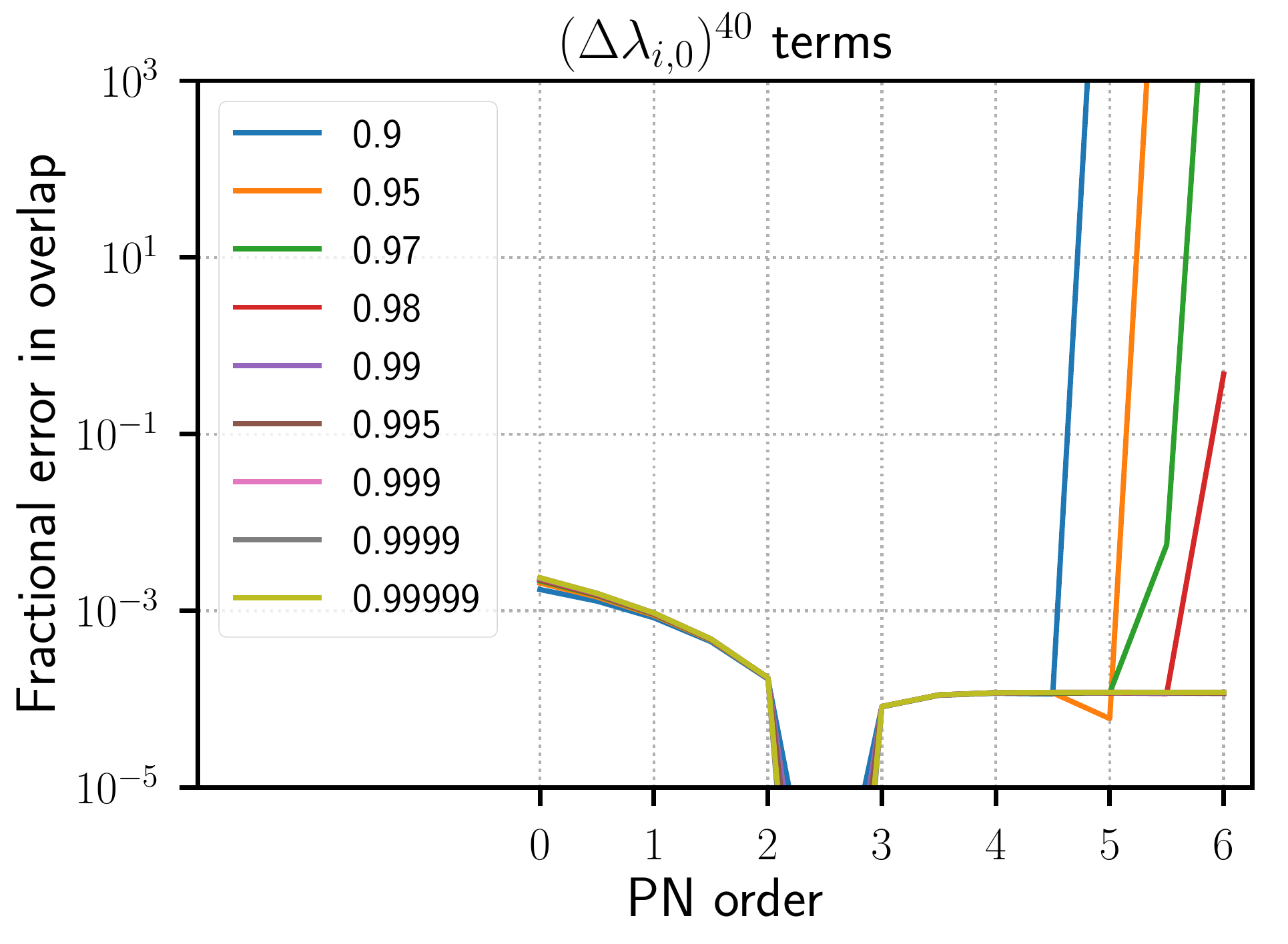}
    \caption{The accuracy at which the analytical approximation
    predicts the overlap between two waveforms
    that differ by a change in a single $\lambda_{i,0}$ term in the
    post-Newtonian expansion. Shown for $\lambda_{i,0}$ terms with corrections up to 4th (top), 20th (middle) and 40th (bottom) order in the Taylor series expansion of the overlap. Note that the Fisher matrix is the 2nd order correction, and all odd-ordered terms are 0.
    An animation of this figure showing the inclusion of
    higher-order corrections one-by-one
 \href{https://icg-gravwaves.github.io/ian_harry/tidal_template_bank/Figure4.mp4}{can be found here.}
    }
    \label{fig:change_one_term_ho}
\end{figure}

We identified previously the assumptions that are made when defining the Fisher-matrix approximation of the overlap.
Now we investigate the validity of the assumption that terms beyond the leading order term in the Taylor series expansion of the overlap are negligible.
We note that there has been some exploration of the importance
of these terms in previous work~\cite{Vallisneri:2007ev, Zanolin:2009mk, Vitale:2010mr, Vallisneri:2011ts, Vitale:2011zx}.
However, these works differ from this one in that they focus
on the bias that neglecting such terms would have on
inferring the parameters of gravitational-wave signals, and do
not consider the specific problem of tidal terms.
If we go back to equation~\ref{eq:match_taylor_expand} and expand the match
to fourth order in $\Delta\theta_i$ we can write the match as
\begin{equation}
\label{eq:match_fourth_order}
 1 - M\left(\theta_i, \Delta\theta_i\right) = - \frac{1}{2!} \frac{\partial^2 M}{\partial \theta^i \partial \theta^j} \Delta \theta_i \Delta \theta_j - P_4
\end{equation}
where $P_4$ is the 4th order correction term given by
\begin{equation}
    P_4 = \frac{1}{4!} \frac{\partial^4 M}{\partial \theta^i \partial \theta^j \partial \theta^k \partial \theta^l} \Delta \theta_i \Delta \theta_j \Delta \theta_k \Delta \theta_l.
\end{equation}
In this case the $P_4$ term involves a rank-4 tensor\footnote{Here we use the term "tensor" as it is used in computational science, rather than in mathematics. Formally this is not a tensor as it will not obey the coordinate transformations rules.}.
If we consider that a standard TaylorF2 waveform might contain contributions from $O(10)$ terms, this tensor must contain $10^4$ terms.
As with the Fisher matrix, there is a lot of symmetry in this tensor, and many terms are repeated.
Our use of $\lambda_{i,j}$ coordinates are again useful here if
we try to evaluate this rank-4 tensor.
In particular we can show that
\begin{multline}
    \frac{\partial^4 M}{\partial \lambda_{i,j} \partial \lambda_{k,l} \partial \lambda_{m,n} \partial \lambda_{o,p}} = \\ - 4 \Re \left[ \int^{f_U}_{f_L} df \frac{f^{(i + k + m + o - 27) / 3} \log^{j+l+n+p} f }{S_h(f)} \right].
\end{multline}

In this way it is also easy to expand this to even higher-order terms in the Taylor series expansion.

To investigate the importance of these terms, we again follow our procedure in Figure~\ref{fig:change_one_term_fisher} where we only perturb one of the PN terms and compute the resulting overlap.
The advantage to doing this is that there is still only one term in the rank-4 tensor that is non-zero, and indeed only one term in all the higher order tensors as well, allowing us to compute analytical overlap to very high order in the Taylor expansion when varying only a single PN order.

As with Figure~\ref{fig:change_one_term_fisher} we numerically determine the perturbation needed to produce a numerical overlap of a given value.
We then compute the analytical overlap up to various orders in the Taylor expansion.
In Figure~\ref{fig:change_one_term_ho} we show results at 4th order, 20th order and 40th order in the Taylor expansion. A short animation perhaps shows this better, demonstrating how the agreement evolves as we incrementally add increasingly higher order terms, \href{https://icg-gravwaves.github.io/ian_harry/tidal_template_bank/Figure4.mp4}{which can be viewed by clicking here.}

We notice that terms at 2.5PN order and below are in most cases
not highly sensitive to the inclusion of the higher-order terms.
Adding the 4th order-term does improve the accuracy of the computation, especially noticeable at smaller values of the 
numerical overlap. The predicted overlap quickly converges though.
It still doesn't converge to an error of 0, hinting again at
the presence of some subtle systematic between the Fisher
matrix and the numerical overlap that we have not accounted for.

At PN orders above 2.5PN order we observe quite different behaviour. It is often the case that the addition of higher-order
terms first causes the accuracy of the overlap calculation to decrease (in many cases predicting overlaps much greater than 1, or much less than 0, which of course aren't possible physically, but are possible from this analytical prediction). However, as increasingly higher-order terms are added, the analytical overlap
does converge to be consistent with the numerical value. For the
6PN term at an overlap of 0.97, this required adding terms up to 60th order in the Taylor series expansion. This indicates that the unreliability of the Fisher matrix predictions \emph{is} due to the missing higher-order terms in the expansion of the analytical overlap computation.

\subsection{What is special about the 2.5PN term}

Our results in Figure~\ref{fig:change_one_term_ho}
beg the question of why the Fisher matrix performs so poorly for the large post-Newtonian terms and why the 2.5 post-Newtonian
term is the one most easily predicted.
To understand this we examine the generic form of the rank-N tensor that would represent the Nth order term in the Taylor expansion.
Considering only the non-log terms $\lambda_{i,0}$, this can be generically written as:
\begin{equation}
   \frac{\partial^2 M}{\partial \lambda_{i} \partial \lambda_{j} \partial \lambda_{k} ...} = 
    4 \Re 
    \left[ (-1)^{N/2} \int^\infty_0 df \frac{f^{(i + j + k + ... - 5N - 7) / 3}}{S_h(f)} \right].
\end{equation}
From examination of this equation we can see that the power
in the numerator of the integral is equal to $(i+j+k+.. - 5N - 7)/3$ or $((i-5) + (j-5) + (k-5) + ... -7)/3$.
Therefore if the sum of the various terms (after subtracting 5 from each) is larger than 7 the power in the numerator of the integral is positive.
At high frequencies $S_h(f)$ is dominated by the shot noise which
is proportional to $f^2$.
Therefore there will be cases where the 
power in the numerator is larger than the $f^2$ term in the
denominator which will make this calculation very sensitive to
content at high frequencies.
It also makes the calculation very sensitive to the \emph{choice}
of the upper frequency cutoff, which we will explore further in
the next subsection.
Even for the Fisher matrix, changes in the 6 PN term result in a numerator term proportional to $f^{7/3}$, which is rising faster
than the shot noise contribution in the denominator.

If one considers the fourth order (and higher) terms in the Taylor series expansion, the power of the numerator can become very large.
Therefore when we move to a higher order, while the $\Delta \theta_i$ terms and the inverse factorial term will become increasing small, the integral will become increasingly large.
In contrast, at 2.5PN order, the numerator of the integral remains $f^{-7/3}$ regardless of what order in the Taylor expansion we are considering.
In this case the higher-order terms in the expansion will quickly be negligible.
Finally, at low PN orders the numerator will have a negative power, which will also grow increasingly large for higher terms in the Taylor series expansion.
Here also these terms can rise faster than the noise floor at low frequencies, although this is somewhat limited by the practical need to start the integral from some low-frequency cutoff.

\subsection{The practicality of including higher-order terms in the match approximation}

We have demonstrated that the Fisher matrix poorly predicts the overlap between nearby waveforms when varying terms at high-post-Newtonian order, as is necessary when considering changes in the tidal deformation terms.
We have also demonstrated that we can accurately predict such overlaps analytically by including higher-order corrections to the Fisher matrix.
However, in our tests we simplified the computation of these higher-order terms by changing only a single post-Newtonian term at a time.

It is much more computationally impractical to compute these higher-order terms analytically when varying multiple post-Newtonian terms, for example if changing the masses of a binary-neutron star waveform with a given equation-of-state.
The reason for this can be seen if we again consider equation~\ref{eq:match_fourth_order}.
Here we include the fourth order $P_4$ term, which might include
$\sim 10^4$ terms for a standard TaylorF2 waveform.
Even after removing duplicated terms due to symmetry we can see that if we include higher-order terms in the analytical computational of the overlap, the computational complexity will grow exponentially with additional terms.
If we include terms to 20th order, which still does not accurately predict matches for tidal post-Newtonian orders, we would require a rank-20 tensor with $\sim 10^{20}$ terms.
\textit{It is simply not practical to compute the analytical match for waveforms including tidal corrections at sufficient order for the match to be reliable.}

\subsection{Sensitivity of the analytical overlap to the upper
frequency cutoff}

In the previous section we have suggested that the validity of the Fisher-matrix approximation for overlap computation is sensitive to the choice of the upper frequency cutoff.
This is due to the fact that the numerator of the Fisher matrix
integral for tidal terms can be as large, or larger, than the
shot noise term from the PSD in the denominator.
We do expect the numerical overlap for TaylorF2 waveforms (when the waveform model does not include a termination condition) to be sensitive to the
choice of upper frequency cutoff, but it is interesting to explore
how both the numerical and Fisher-matrix approximated overlap
computations vary based on the choice of termination frequency.

\begin{figure}[tp]
    \includegraphics[width=\columnwidth]{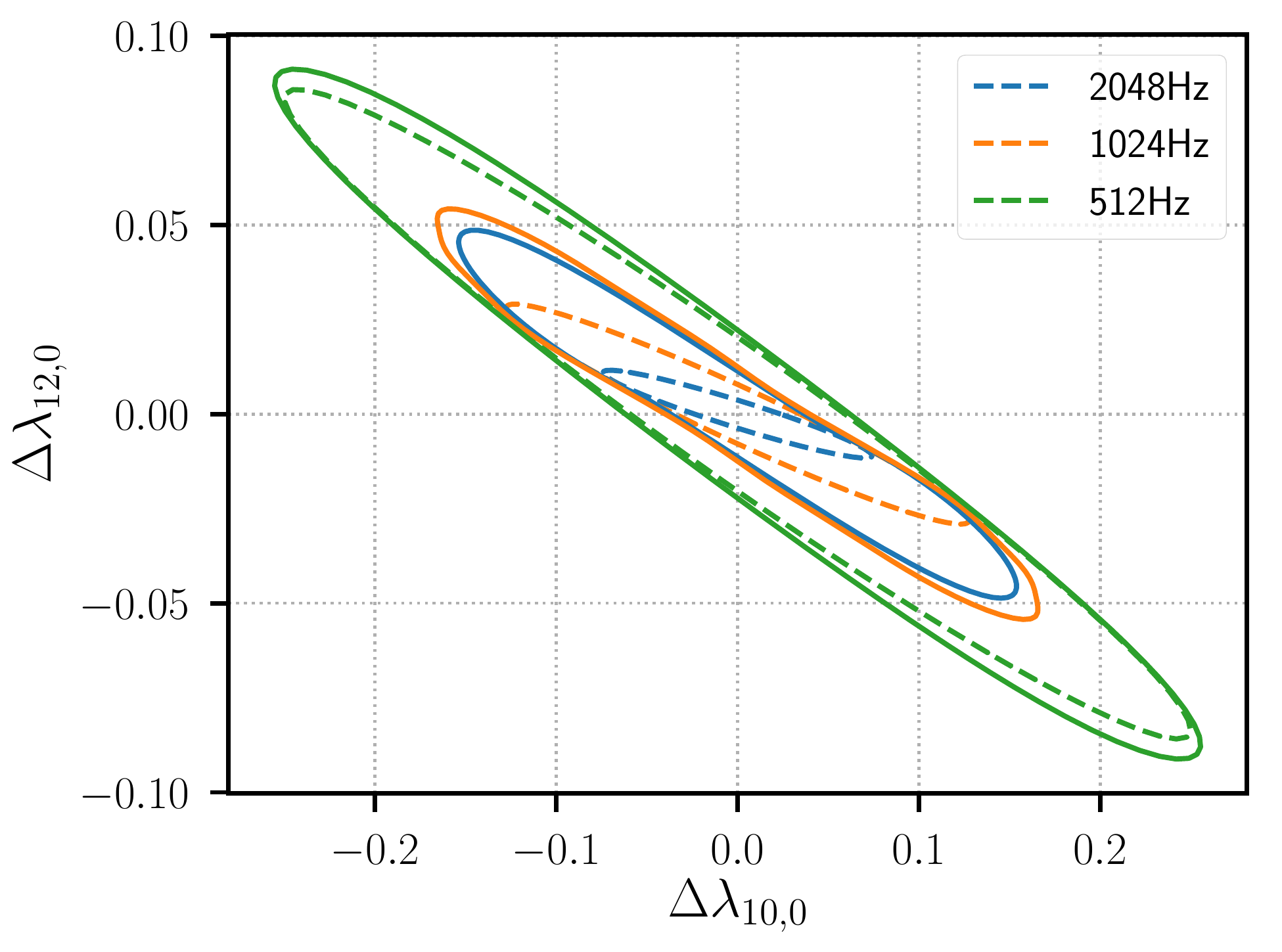}
    \caption{
    The perturbation that needs to be added to the 5- and
    6-PN terms to achieve a overlap of 0.97 with an unperturbed
    TaylorF2 waveform as a function of termination frequency.
    Solid lines shows the numerical overlap.
    Dotted lines show the overlap predicted by the Fisher matrix.
    An animated version showing how this figure changes as a function of the termination frequency and as a function of the overlap
    \href{https://icg-gravwaves.github.io/ian_harry/tidal_template_bank/Figure5.mp4}{ can be found here.}
    }
    \label{fig:change_flower}
\end{figure}

To illustrate this we reproduce Figure~\ref{fig:tidal_metric},
where we identified the perturbation in the 5-PN and 6-PN terms that
is necessary to give an overlap of 0.97 with an unperturbed waveform.
We compute this, as before, for both numerical and analytical overlap
computations, but now compute this at 3 values of the termination
frequency, 512Hz, 1024Hz and 2048Hz.
The results of this are shown in Figure~\ref{fig:change_flower}.

From this we observe the dependence of the numerical overlap of the
TaylorF2 waveform on this choice. There is a significant difference
between 512Hz and 1024Hz, but not much of a change when increasing
further to 2048Hz.
In contrast, the accuracy of the Fisher matrix prediction is relatively good at 512Hz, but decreases rapidly as the termination frequency continues to increase.

\subsection{Discussion}

The main result of this section is that the Fisher Matrix-derived
metric is not
suitable for predicting overlaps between two waveforms modelled
by the analytical post-Newtonian TaylorF2 approximation.
The approximation that the parameter space can be described by a
Riemannian metric given by the Fisher Matrix breaks down when
considering TaylorF2 waveforms including tidal terms.
We have demonstrated under which conditions one would obtain
inaccurate overlaps and shown that including higher-order terms
in the expansion of the expression for the overlap does resolve
the problem, but is computationally impractical for realistic use cases.

While we can look to mitigate this issue by carefully choosing the
upper frequency cutoff to use, it is possible that this effect is
particular to the waveform approximation used~\footnote{We acknowledge the anonymous referee for their suggestion to explicitly include a paragraph of this nature}. In particular the
TaylorF2 waveform is modelled as a series of post-Newtonian terms,
dependent on increasing powers of frequency, that cutoff abruptly
at some termination condition. As discussed above, this can lead
the derivatives used in the analytical overlap computation to be
highly sensitive to the choice of termination frequency.
It is possible that complete waveforms including inspiral, merger
and ringdown, which do not terminate abruptly, will behave better
in this regard.
However, there does not exist a parameterization for such waveforms
that will allow us to construct a globally-flat metric on the
parameter space, which is required for many of the studies we
perform here. Therefore we leave an exploration of these effects
with alternative waveform models for future work.

\section{Using stochastic placement to create a tidal template bank}

The focus of this work has been on exploring the inadequacies
of the Fisher matrix approximation in accurately predicting the overlap between
two waveforms due to differences in the tidal PN terms.
However, the original motivation for exploring this was the goal of being able to create template banks of filter waveforms that include tidal corrections.
We have already argued that geometric placement will not be appropriate here, which will also render the hybrid method proposed in~\cite{Roy:2017qgg, Roy:2017oul} inappropriate here as well.
However, stochastic template placement~\cite{Harry:2008yn, Babak:2008rb, Harry:2009ea, Manca:2009xw, Ajith:2012mn, Privitera:2013xza, Capano:2016dsf} can be used when computing overlaps numerically and therefore remains an appropriate solution to this problem.

To demonstrate this we create a template bank of waveforms to cover systems with both component masses $\in [1,3] M_{\odot}$, with both component dimensionless spin magnitudes $\in [-0.05,0.05]$ and with the tidal deformability $\lambda$ parameter~\cite{Flanagan:2007ix} for both bodies $\in [0,2000]$.
We also assume the Advanced LIGO zero-detuned, high-power noise curve~\cite{TheLIGOScientific:2014jea}.
This bank contains $69250$ templates, compared to a template bank generated in the same manner, except assuming the tidal deformability for both bodies is 0, which has $41439$ templates.
For comparison, a template bank generated with the geometric algorithm for the same parameter space, ignoring tidal deformability, contains 37977 templates.

We acknowledge that this parameter space is not particularly well motivated physically. Even if neutron stars with masses up to $3 M_{\odot}$ are possible, the tidal deformability parameter would be much smaller for such high-mass stars~\cite{LIGOScientific:2019eut}. We can also reasonably assume that both neutron stars must be governed by the same underlying physics, such that there must be a tight correlation between the two stars' values of $\lambda$ given both masses. Including this physics would considerably reduce the number of additional templates required to create a tidal template bank. The techniques already exist to do this, as used in~\cite{Pannarale:2014rea} for the case of targeting compact binary mergers that might power GRBs. We leave the choice of what exact constraints to use here open, as these constraints will likely evolve rapidly with future observations of binary neutron star mergers.  

\section{Conclusion}

In this paper we have explored the validity of the Fisher matrix
for predicting waveform overlaps, with a particular focus on applications to template bank placement.
We have found that the accuracy of the Fisher matrix-predicted overlaps is poor when considering changes in post-Newtonian terms
larger than ``4PN'' order.
We have investigated the reason for these poor predictions and identify that the neglected higher-order terms in the Taylor series expansion of the overlap are crucial for accurately predicting overlaps at high post-Newtonian order when using the TaylorF2 waveform model.
Unfortunately, the computational cost of including such higher-order terms in analytic predictions quickly becomes prohibitive (indeed it quickly becomes more expensive than a brute-force computation of the overlap).
We therefore conclude that the current geometric template bank placement algorithms are not suitable for placing template banks of binary neutron mergers including tidal deformability modelled using the TaylorF2 waveforms.
We recommend that stochastic placement, using numerical computation of the match, be used instead.

\begin{acknowledgments}

The authors would like to thank the anonymous referees for helpful feedback and critique on the first version of this manuscript.
The authors would like to thank Bhooshan Gadre for useful discussion, comments and feedback on this work. IH thanks the STFC for support through the grant ST/T000333/1. AL thanks the STFC for support through the grant ST/S000550/1.

\end{acknowledgments}

\bibliography{biblio}

\end{document}